
\baselineskip=18pt plus 3pt
%
%
\def\eq#1\endeq
{$$\eqalignno{#1}$$}
\def\leq#1\endeq
{$$\leqalignno{#1}$$}
%
%
\def\qbox#1{\quad\hbox{#1}\quad}
%
%

%
%
\def\nbox#1{\noalign{\hbox{#1}}}
%
%

%
%
\def\proof#1\par{\medbreak\noindent{\it Proof.\/}\quad{#1}\par
    \ifdim\lastskip<\medskipamount\removelastskip\penalty55\medskip\fi}
%
%
\def\remark#1\par{\medbreak\noindent{\it Remark.}\quad#1\par
    \ifdim\lastskip<\medskipamount\removelastskip\penalty55\medskip\fi}
\outer\def\Beginsection#1#2\par{\vskip0pt plus.3\vsize\penalty-250
  \vskip0pt plus-.3\vsize\bigskip\vskip\parskip
  \message{#1}\noindent\hangindent=#2 \hangafter=1
  {\bf#1}\smallskip\noindent}
%
\input mssymb
\font\germ=eufm10
\def\goth#1{\hbox{\germ#1}}
\def\g{{\goth g}}
\def\h{{\goth h}}
\def\A{{\goth A}}
\def\C{{\Bbb C}}
\def\Z{{\Bbb Z}}
\def\Sl{{\it Sl}}
\def\sl{{\it sl}}
\def\al{\alpha}
\def\del{\delta}
\def\La{\Lambda}
\def\la{\lambda}
\def\pa{\partial}
\def\Ga{\Gamma}
\def\op{\oplus}
\def\ot{\otimes}
\def\ome{\omega}
\def\Del{\Delta}
\def\tB{\tilde B}

\def\bB{\bar B}
\def\bn{\bar \nu}
\def\bY{\bar Y}
\def\hA{\hat \A}
\def\hg{\hat \g}
\def\gtg{G\times G}
\def\gog{\g\oplus \g}
\def\cb{C_B(G)}
\def\cbl{C_B^\lambda(G)}
\def\ap{{\mathop a\limits^+}}
\def\bp{{\mathop b\limits^+}}
\def\oc{\mathop \oplus\limits_\chi}
\magnification=\magstep1
\font\twelvebf=cmbx12
\vglue 2cm
\centerline{\twelvebf Regular representations of affine
Kac-Moody algebras}
\bigskip
\centerline{B. Feigin, S. Parkhomenko}
\vskip2cm
\beginsection
\S 1. Introduction

In this paper we investigate one Wakimoto-type construction
of affine Kac-Moody algebras. Our aim is to obtain a version
of the regular representation of the current algebras.

Let us discuss first what is the regular representation for
finite-dimensional complex semisimple Lie groups. Let $G$
be such a group and $C(G)$ be the space of algebraic functions
on $G$. It is clear that $C(G)$ is a $\gtg$-module where $G$
acts from the left and from the right;
the formula for the action is:
$
(g_1, g_2)u=g_1\cdot u\cdot g_2^{-1}, g_1, g_2, u\in G.
$
It is well-known that $C(G)$ as a $\gtg$-module is
$\op V\ot V^*$ where the sum goes over all irreducible
finite-dimensional representations of $G$. Usually
$C(G)$ is called the regular representation. Now let us
fix some Borel subgroup $B\subset G$ and denote by
$\cb$ the space of distributions on $G$ with support on
$B$ and which are smooth along $B$. In other words
$\cb$ is isomorphic to the space of the local
cohomologies of the sheaf of functions on $G$ with support
on $B$, they are non-trivial only in dimension
$\dim G - \dim B$. The group $\gtg$ does not act on
$\cb$, but the Lie algebra $\gog$ of the group $\gtg$
does. Let $b$ be the Lie algebra of the Borel subgroup
$B$. It is evident that the restriction of the
$\gog$-module $\cb$ to $b\op b$ can be integrated to the
representation of $B\times B$. It means that $\cb$
belongs to the ``category of representations with highest
weight''. In $\cb$ we have the distinguished subspace
$\del(B)$, which is invariant with respect to $B\times B$
and isomorphic to the space of sections of the line bundle
$\xi$ on $B$, $\xi$ is $\La^NT$, where $T$ is the normal
bundle to $B$ in $G$ and $N$ is the dimension of $T$.
It is easy to find all $n\op n$-invariant vectors in
$\del(B)$, where $n$ is the maximal nilpotent subalgebra
in $b$. They are labeled by the elements from the weight
lattice $\Ga$ and the weight of such vector
$v(\chi), \chi\in \Ga$ with respect to the sum $\h\op \h$
of two Cartan sub-algebras is $(\chi, -2\rho-\chi)$,
where $2\rho$ is a sum of all positive roots of $\g$.

This construction can be generalized by the following way.
Let us consider the formal vicinity $\tB$ of the
submanifold $B$ in $G$. The fundamental group
$\pi_1(\tB)\cong \pi_1(B)\cong \pi_1(H)$, where $H$ is
maximal torus in $G$. Therefore, each element
$\la\in \h^*$ defines a one-dimensional bundle $\nu_\la$
on $\tB$ with flat connection and this connection is
uniquily determined by the restriction of $\nu_\la$ on
 $H$. Lie algebra $\g\op \g$ acts in the space of sections
of $\nu_\la$ and also on the space of local cohomologies
of $\nu_\la$ with the support on $B$. We denote this
space by $\cbl$. Again in $\cbl$ it is possible to
construct the set of vacuum vectors $V(\chi), \chi\in \Ga$
such that the character of $\h\op \h$ is a pair
$(\chi+\la, -\la-\chi-2\rho)$. It gives us for generic
$\la\in\h^* a$ $ \g\op\g$ module $\cbl$ which is isomorphic to
the sum $M_{\chi+\la}\ot M_{-\chi-\la-2\rho}, \chi\in \Ga, M_u$
is a Verma representation with highest weight $u$. So this
is an analog of the decomposition $C(G)= \op V\ot V^*$.
Note, that the algebra $C(G)$ acts in $\cbl$, each function
$f\in C(G)$ defines an operator $\cbl\to \cbl$ which is
just the multiplication on $f$. Therefore we obtain some
``vertex operator'':
$
(V\ot V^*)\ot(\oc M_{\chi+\la}\ot M_{-\chi-\la-2\rho})
\to\oc M_{\chi+\la}\ot M_{-\chi-\la-2\rho}, V
$
is a finite-dimensional representation of $\g$.
The important thing is that this algebra of vertex operators
is commutative.

Our aim now is to provide the same construction for infinite-
dimensional Lie algebras. To do it let us recall some main
ideas of constructing of Wakimoto representations.
Let $M$ be some (may be infinite-dimensional) manifold and
Lie $(M)$ be the Lie algebra of vector fields on $M$.
Natural representations of Lie $(M)$ are realized in
different spaces of distributions on $M$. Let us denote by
$C_N(M)$ the space of distributions on $M$ with support on
$N$ and which are smooth along $N$. Another name for this
space -- local cohomologies of the sheaf of functions with
support on $N$. If $M$ is the finite dimensional manifold
then $C_N(M)$ is a representation of Lie $(M)$, but in the
infinite-dimensional case the situation is more complicated.
The machinery of local cohomologies does not work in the
infinite dimensional case, so we have to construct $C_N(M)$
by hands and then again by hands we have to verify the
functorial properties of $C_N(M)$. In other words the
construction of $C_N(M)$ depends on the choice of the
coordinate system in the small neighbourhood of $N$ and
we need to know what it will be if we change the coordinates.
Infinitesimally we want to determine the action of
Lie $(M)$ on $C_N(M)$.

The construction of $C_N(M)$ is the following. For simplicity
we suppose that $N$ is isomorphic to the affine space and let
us fix the coordinate system:
$\{ x_1, x_2,\cdots, y_1, y_2,\cdots\}$
in the neighbourhood of $N$ such that $\{ y_1, y_2,\cdots\}$
are the coordinate in the ``normal direction to $N$'',
it means that all $y_j$ are zero on $N$ and they constitute the
coordinate system on the transversal to $N$ submanifold.
The functions $\{ x_i\}$ form coordinates on $N$. Now let
$D$ be an algebra of differencial operators with generators:
$\{ x_i, y_j, \pa/\pa x_i, \pa/\pa x_j\}$.
We define $C_N(M)$ as an irreducible representation of $D$ with
the vacuum vector vac, such that $y_j{\rm vac}=0,\ j=1,2,\cdots$
and $\pa/\pa x_i {\rm vac}=0,\ i=1,2,\cdots$. In [2,4] it is
shown that on $C_N(M)$ the central extension of Lie $(M)$
by Lie $(M)$-module $C(M)$ acts, where $C(M)$ is an algebra of
functions on $M$. This extension is nontrivial if
$\dim(N)={\rm codim}(N)=\infty$. This means that the infinitesimal
group of symmetries of local cohomologies is the Lie algebra
of twisted differential operators on the manifold of the order
$\le 1$. The general theory of such cohomologies should exist, as in
finite-dimensional case, with suitable modifications.
We are interested in the following particular case of the
main construction. Let $A$ be a semisimple complex Lie group,
$\A$-Lie algebra of $A, X$ be a homogeneous space of $A$,
$LA$ be the loop group of $A$, $LA$ consists of maps
$S^1\to A$, and $LX$ be the space of maps $S^1\to X$.
It is clear that we have the map $L\A\to {\rm Lie} (LX)$,
where $L\A$ is the Lie algebra of the group $LA$.
Choose the submanifold $N\subset LX$, which consists of
boundary values of the analitic maps from the disk
$|z|\le 1$ into $X$. In the space $C_N(LX)$ the central
extension of the Lie algebra $L\A$ by the space of functions
$C(LX)$ acts. In some cases this extension can be
transformed (by the adding of a coboundary) into the
extension with values only in the constants $\C\subset C(LX)$.
Therefore in this cases we obtain a representation of the
affine Lie algebra of $L\A$ in the space $C_N(LX)$.

We know at least two situations where it is possible to make
the reduction of the extension to the constants. The first one
is the case $X=A/B, B$ is the Borel subgroup in $A, X$ is the
flag manifold for $A$. The affine algebra $\hA$ acts in
$C_N(LX)$ with level $-g$, where $g$ is the dual Coxeter
number for $A$. The slight variation of this construction
gives us Wakimoto modules of arbitrary level [2].
The second example is when $A=G\times G$ and $X=A/G_\Del$, where $G_\Del$
is a diagonal subgroup in $G\otimes G$. In other words, $LX$ is
the loop group $LG$, where $LA=LG\oplus LG$ acts -- one
$LG$ by left shifts and the other by the right shifts.
So we want to define the ``regular'' representation in the
space of distributions on $LG$. It is possible to prove that
we obtain the $\hg\op \hg$-module $C_N(LX)$, where the
right and the left algebras $\hg$ act with central charge
$-g$. Note, that the space $C_N(LX)=C_N(LG)$ formally is
very close to the space $C_B(G)$ which we discussed in the
first part of the introduction. Actually, $N$ is a parabolic
subgroup in $LG$, but we certainly can replace $N$ by the
Borel subgroup $\bB$ in the loop group $LG$. The
corresponding space $C_{\bB}(LG)$ is also easy to define
and the values of the central charges will be the same.
It is also possible to construct the family of representations
$C_{\bB}^\la(LG)$, where $\la$ is the character of the
Cartan sub-algebra of $\g$. But if we want to change the values
of the central charges we need a more subtle construction.

Suppose, the Lie group $H$ is acting on its homogeneous space
$Y, H_0$ is stationary subgroup of the point, $C(Y)$- the
space of functions on $Y, h$ and $h_0$ are the Lie algebras of
$H$ and $H_0$. The first way to deform the action of $h$ on
$C(Y)$ is the following. Fix a 1-cocycle $\ome:h\to C(Y)$ and
add this cocycle to the action of $h$. It means that the new
action of $u\in h$ is given by the formula
$u_\ome(f)=u(f)+\ome(u)\cdot f$. Because of the cocycle
condition it gives us a representation $h\to {\rm End} \bigl(C(Y)\bigr)$.
The similar thing is true for the group action.
Now suppose that $\nu: \La^2(h)\to \C$ is 2-cocycle and
$\mu: h\to C(Y)$ is 1-cochain such that
$d\mu(u_1, u_2)\in \C\cdot 1\subset\C(Y)\ u_1, u_2\in h$
so $d\mu$ determine 2-cochain with values in $\C$ and we
suppose that $d\mu=\nu$. In this case the formula
$u_\mu(f)=u(f)+\mu(u)\cdot f, u\in h$ gives us a
projective representation of $h$ and the corresponding
2-cocycle is $\nu$. If we work with the action of groups,
then $H^2(H, C(Y))\cong H^2(H_0, \C),$ so if we fix a
class $\bn\in H^2(H, \C)$ such that the image of $\bn$ in a map
$H^2(H, \C)\to H^2(H_0, \C)$ is zero, then we can construct
the projective action of $H$ in the $C(Y)$ such that the
corresponding cocycle in $\bn$. Infinitesimal version of
it is also true, if we replace $Y$ by its contractible subspace
$\bY$.

Now let us apply these arguments to the case,
when $H=LG\times LG$
and $Y=LG$.
Let $\bar Y$ be some contractible open set in $Y$.
Using the arguments with cocycles it is possible
to show that on $C(\bar Y)$ there exists
the projective representation of
$L\g\oplus L\g$ by the differencial operators
of degree $\le 1$ such that the corresponding
2-cocycles are $(m\omega,-m\omega)$, $m\in\C$
and $\omega$ is the standard 2-cocycle of
$L\g$.
In the space of distribution
$C_N(Y)$ we get the representation of
$\hat\g\oplus\hat\g$ and the levels are
$(m-g,-m-g)$.
The diagonal subalgebra
$\hat\g_\Delta\subset\hat\g\oplus\hat\g$ has level
$-2g$.
This is the case when we can add ghosts and compute semi-infinite
homology which are the candidate for the space of fields in this version
of topological field theory.
\par
The paper is arranged as follows.
\par
\noindent
In Sect.\ 2 we briefly discuss the construction of regular
representations in the simplest
$G=\Sl(2,\C)$ and $G=\Sl(3,\C)$ cases.
In these examples the submanifolds $N$ are chosen by using
the Gauss decompositions of $G$.
In Sect.\ 3 we give the loop versions of constructions of Section 2.
As the space of distributions
$D(N,LG,\xi_m)$ we consider a Fock module generated from
a vacuum vector by a spin $(1,0)$ conjugated bosonic fields
and the submanifold $N$ is a set of boundary values
of analytic maps from the unit disk
into the open subset of $G$, which is defined by the Gauss decomposition
of $G$. In this situation $m$
has arbitrary value.
Sect.\ 4 is devoted to the generalization on
$\widehat{\sl(n+1,\C)}$ case.
We think that our representations can be used as an ingredient of
$G/G$ topological field theory.
\beginsection
\S2. Regular representation in the finite-dimensional case

Let us briefly discuss the regular representations
of $\sl(2,\C)$ and $\sl(3,\C)$ Lie algebras.
\par
Using the Gauss decomposition let us to introduce coordinate
systems in the open subsets
$\overline{\Sl(2,\C)}$ and
$\overline{\Sl(3,\C)}$ of $\Sl(2,\C)$ and $\Sl(3,\C)$:
$x\in \overline{\Sl(2,\C)}$ if $x$ can be represented as a product
$$
x=\left[\matrix{
1&0\cr
x_1&1\cr}
\right]
\left[\matrix{
\exp(-y)&0\cr
0&\exp(y)\cr}
\right]
\left[\matrix{
1&x^1\cr
0&1\cr}
\right];
\eqno(1)
$$
similary $x\in\overline{\Sl(3,\C)}$ if
$$
x=\left[\matrix{
1&0&0\cr
x_1&1&0\cr
x_3&x_2&1\cr}
\right]
\left[\matrix{
\exp(-y_1)&0&0\cr
0&\exp(y_1-y_2)&0\cr
0&0&\exp(y_2)\cr}
\right]
\left[\matrix{
1&x^1&x^3\cr
0&1&x^2\cr
0&0&1\cr}
\right].
\eqno(2)
$$
We shall call them Gauss coordinate systems.
Let $E,H,F$ be the standard generators of
$\sl(2,\C)$ and $E_i,H_i,F_i$, $i=1,2$
be the standard generators of
$\sl(3,\C)$.
The following formulas give us the left and right actions
of $\sl(2,\C)$ and  $\sl(3,\C)$ in the Gauss coordinates.
Symbols $L$ and $R$ will be used for the left right actions,
respectively.
\eq
L_E=&{\pa\over\pa x_1};\quad
R_E=\exp(2y){\pa\over\pa x_1}-x^1{\pa\over\pa y}-(x^1)^2{\pa\over\pa x^1}\cr
L_H=&-2x_1{\pa\over\pa x_1}-{\pa\over\pa y};
\quad
R_H={\pa\over\pa y}+2x^1{\pa\over\pa x^1}&(3)\cr
L_F=&-(x_1)^2{\pa\over\pa x_1}-x_1{\pa\over\pa y}+\exp(2y){\pa\over\pa x^1};
\quad
R_F={\pa\over\pa x^1}\cr
L_{E_1}=&{\pa\over\pa x_1};~
R_{E_1}=\exp(2y_1-y_2){\pa\over\pa x_1}+
\exp(2y_1-y_2)x_2{\pa\over\pa x_3}\cr
&-x^1{\pa\over\pa y_1}-x^1x^3{\pa\over\pa x^3}+
(x^1x^2-x^3){\pa\over\pa x^2}-(x^1)^2{\pa\over\pa x^1}\cr
L_{E_2}=&{\pa\over\pa x_2}+x_1{\pa\over\pa x_3};~
R_{E_2}=\exp(-y_1+2y_2){\pa\over\pa x_2}
-x^2{\pa\over\pa y^2}
-(x^2)^2{\pa\over\pa x^2}
+x^3{\pa\over\pa x^1}\cr
L_{H_1}=&-2x_1{\pa\over\pa x_1}+x_2{\pa\over\pa x_2}
-x_3{\pa\over\pa x_3}-{\pa \over\pa y_1};\cr
&R_{H_1}={\pa\over\pa y_1}+x^3{\pa\over\pa x^3}
-x^2{\pa\over\pa x^2}+2x^1{\pa\over\pa x^1}&(4)\cr
L_{H_2}=&x_1{\pa\over\pa x_1}-2x_2{\pa\over\pa x_2}
-x_3{\pa\over\pa x_3}-{\pa\over\pa y_2};~
R_{H_2}={\pa\over\pa y_2}+x^3{\pa\over\pa x^3}
+2x^2{\pa\over\pa x^2}-x^1{\pa \over\pa x^1}\cr
L_{F_2}=&x_3{\pa\over\pa x_1}-(x_2)^2{\pa\over\pa x_2}-x_2{\pa\over\pa y_1}
+\exp(-y_1+2y_2){\pa\over\pa x_2};~
R_{F_2}=x^1{\pa\over\pa x^3}+{\pa\over\pa x^2}\cr
L_{F_1}=&-(x_1)^2{\pa\over\pa x_1}
+(x_1x_2-x_3){\pa\over\pa x_2}
-x_1x_3{\pa\over\pa x_3}
-x_1{\pa\over\pa y_1}\cr
&+\exp(2y_1-y_2)x^2{\pa\over\pa x^3}
+\exp(2y_1-y_2){\pa\over\pa x^1};~
R_{F_1}={\pa\over\pa x^1}\cr
\endeq
The formulas (3), (4) give embeddings of the Lie algebras
$\sl(2,\C)\oplus\sl(2,\C)$
and $\sl(3,\C)\oplus\sl(3,\C)$
into the Lie algebras of vector fields
on $\overline{\Sl(2,\C)}$ and $\overline{\Sl(3,\C)}$,
and equip the spaces of the distributions
$D(\overline{\Sl(2,\C)},\Sl(2,\C))$ and
$D(\overline{\Sl(3,\C)},\Sl(3,\C))$
with structures of modules over those Lie algebras.
The embeddings (3), (4) will be called the regular representations.
Regular representations of other finite dimensional Lie algebras
may be derived in a similar manner.
\beginsection
\S3. Regular representations of $\widehat{\sl(2,C)}$ and
$\widehat{\sl(3,C)}$ affine Kac-Moody algebras

In this section we state affine analogues of the formulas
(3), (4).
Let us consider the simplest case of $\widehat{\sl(2,\C)}$
algebra.
Let $a_1(z), \ap_1(z), a^1(z),\ap^1(z),
b(z), \bp(z)$
be three conjugate pairs of bosonic fields of
spin (1,0) with the usual operator expansions:
\eq
&a_1(z)\ap_1(w)=a^1(z)\ap^1(w)=b(z)\bp(w)=(z-w)^{-1}+\cdots\cr
&a_1(z)=\sum_{n\in\Z}a_1(n)z^{-n-1};~
a^1(z)=\sum_{n\in\Z}a^1(n)z^{-n-1};~
b(z)=\sum_{n\in\Z}b(n)z^{-n-1}&(5)\cr
&\ap_1(z)=\sum_{n\in\Z}a_1(n)z^{-n};~
\ap^1(z)=\sum_{n\in\Z}\ap^1(n)z^{-n};~
\bp(z)=\sum_{n\in\Z}\bp(n)z^{-n}\cr
\endeq
The fields $a^1,a_1,b$ are the loop algebra versions of the operators
${\pa\over\pa x^1},{\pa\over\pa x_1},{\pa\over\pa y}$
and the fields
$\ap^1,\ap_1,\bp$
are the loop algebra versions of the operators
$x^1,x_1,y$.
Let $\Gamma$ be the Heisenberg algebra,
generated by $a_1,\ap_1,a^1,\ap^1,b,\bp$
and $M$ be the irreducible representation of
$\Gamma$ with the vacuum vector, annihilated by
$\ap_1(n),\ap^1(n),\bp(n),n>0$,
and by $a_1(n),a^1(n),b(n),n\ge 0$.
$M$ can be identified with some space of distributions
on the manifold $L\overline{\Sl(2,\C)}$
of loops on $\overline{\Sl(2,\C)}$.
The regular representation of $\widehat{\sl(2,\C)}\oplus
\widehat{\sl(2,\C)}$ is given by the formulas:
\eq
&L_E=a_1;~R_E=\exp(2\bp)a_1-\ap^1b-:
(\ap^1)^2a^1:-(k+4)\pa\ap^1
+(k+2)\ap^1\pa\bp\cr
&L_H=
-2:\ap_1a_1:-b-(k+2)\pa\bp;~
R_H=b+2:\ap^1a^1:-(k+2)\pa\bp&(6)\cr
&L_F=-:(\ap_1)^2a_1:-\ap_1b+\exp(2\bp)a^1+k\pa\ap_1-
(k+2)\ap_1\pa\bp;~
R_F=a^1\cr
\endeq
\par
The regular representation of
$\widehat{\sl(3,\C)}$  algebra is given in a similar way.
Let $a_i(z),\ap_i(z),$\hfill \break
$a^i(z),\ap^i(z)$, $i=1,2,3$,
$b_i(z),\bp_i(z)$, $i=1,2$
be a set of spin (1,0) conjugate pairs of bosonic fields
with operator expansions:
\eq
&a_i(z)\ap_j(w)=a^i(z)\ap^j(w)=(z-w)^{-1}\delta_{ij}+\cdots,~
i,j=1,2,3&(7)\cr
&b_i(z)\bp_j(w)=(z-w)^{-1}\delta_{ij}+\cdots,~
i,j=1,2\cr
\endeq
Then
\eq
L_{E_1}=&a_1\cr
R_{E_1}=&(a_1+\ap_2a_3)\exp(2\bp_1-\bp_2)
-\ap^1b_1-:\ap^1\ap^3a^3:+:
(\ap^1\ap^2-\ap^3)a^2:\cr
&-:(\ap^1)^2a^1:-\ap^1
(-(k+3)\pa\bp_1+\alpha_1\pa\bp_2)-(k+6)\pa\ap^1\cr
L_{E_2}=&a_2+\ap_1a_3\cr
R_{E_2}=&a_2\exp(-\bp_1+2\bp_2)-\ap^2b_2-:(\ap^2)^2a^2:+
\ap^3a^1\cr
&-\ap^2(\alpha_2\pa\bp_1-(k+3)\pa\bp_2)
-(k+5)\pa\ap^2\cr
L_{H_1}=&-2:\ap_1a_1:+:\ap_2a_2:-:\ap_3a_3:-b_1-(k+3)
\pa\bp_1+\alpha_2\pa\bp_2&(8)\cr
R_{H_1}=&b_1+:\ap^3a^3:-:\ap^2a^2:+2:\ap^1a^1:-(k+3)\pa\bp_1
+\alpha_1\pa\bp_2\cr
L_{H_2}=&:\ap_1a_1:-2:\ap_2a_2:-:\ap_3a_3:-b_2+\alpha_1\pa\bp_1
-(k+3)\pa\bp_2\cr
R_{H_2}=&b_2+:\ap^3a^3:+2:\ap^2a^2
:-:\ap^1a^1:+\alpha_2\pa\bp_1-(k+3)\pa\bp_2\cr
L_{F_2}=&\ap_3a_1-:(\ap_2)^2a_2:-
\ap_2b_2+\exp(-\bp_1+2\bp_2)a^2\cr
&+\ap_2(\alpha_1\pa\bp_1-
(k+3)\pa\bp_2)+(k+1)\pa\ap_2\cr
R_{F_2}=&\ap^1a^3+a^2\cr
L_{F_1}=&-:(\ap_1)^2a_1:+:(\ap_1\ap_2-\ap_3)a_2:
-\ap_1:\ap_3a_3:-\ap_1b_1\cr
&+(a^1+\ap^2a^3)
\exp(2b_1-b_2)+\ap_1(-(k+3)\pa\bp_1+
\alpha_2\pa\bp_2)+k\pa\ap_1\cr
R_{F_1}=&a^1,\cr
\endeq
where $\alpha_1,\alpha_2$ are complex numbers such that
\eq
&\alpha_1+\alpha_2=k+3&(9)\cr
\endeq
\par
In contrast to the  $\widehat{\sl(2,\C)}$
case, formulas (8) depend on one arbitrary value
$\alpha=\alpha_1-\alpha_2$.
But there is the change of variables
$b_1(z), b_2(z)$
preserving the relations (7)
and eliminating the dependence on
$\alpha$ in (8):
\eq
&b_1(z)\to b_1(z)-\alpha_2\pa\bp_2(z),&(10)\cr
&b_2(z)\to b_2(z)+\alpha_2\pa\bp_1(z).\cr
\endeq
In conclusion of this section we introduce another form
of representations
$\widehat{\sl(2,\C)}$ and
$\widehat{\sl(3,\C)}$ algebras
which seems important in connection with Wakimoto representations.
Let us introduce a free bosonic fields by equations:
\eq
\pa\rho=&{i\over\sqrt{k+2}}((k+2)\pa\bp-b);~
\pa\lambda={i\over\sqrt{k+2}}((k+2)\pa\bp+b)&(11)\cr
\pa\rho_1=&{i\over\sqrt{k+3}}((k+3)\pa\bp_1-\alpha_1\pa\bp_2-b_1);\cr
&\pa\lambda_1={i\over\sqrt{k+3}}((k+3)\pa\bp_1-\alpha_2\pa\bp_2+b_1)&(12)\cr
\pa\rho_2=&{i\over\sqrt{k+3}}((k+3)\pa\bp_2-\alpha_2\pa\bp_1-b_2);\cr
&\pa\lambda_2={i\over\sqrt{k+3}}((k+3)\pa\bp_2-\alpha_1\pa\bp_1+b_2).\cr
\endeq
The operator product expansions of fields (11), (12)
are given by:
\eq
&\rho(z)\rho(w)=-\lambda(z)\lambda(w)=2ln(z-w)+\cdots,&(13)\cr
&\rho_i(z)\rho_j(w)=-\lambda_i(z)\lambda_j(w)=K_{ij}
ln(z-w)+\cdots,&(14)\cr
\endeq
where $K_{ij}$ is the Cartan matrix of $\sl(3,\C)$.
Then $\widehat{\sl(2,\C)}\oplus \widehat{\sl(2,\C)}$
currents are given by
\eq
L_E=&a_1; R_E=-:({\mathop a \limits^+}^1)^2a^1:-(k+4)\pa
{\mathop a\limits^+}^1-i\sqrt{k+2}{\mathop a\limits^+}^1
\pa \rho +\exp [-{i\over {\sqrt{k+2}}}(\rho +\lambda )]a_1\cr
L_H=&-2:{\mathop a\limits^+}_1a_1:+i\sqrt{k+2} \pa \lambda;
R_H=2:{\mathop a\limits ^+}^1a^1:+i\sqrt{k+2} \pa \rho &(15)\cr
L_F=&-:({\mathop a\limits^+}_1)^2a_1:+k\pa {\mathop a\limits^+}_1
+i\sqrt{k+2} {\mathop a\limits^+}_1\pa \lambda +\exp
[-{i\over{\sqrt{k+2}}}(\rho +\lambda)]a^1; R_F=a^1\cr
\endeq
{\it Remark.}
We see that these formulas are very close to the standard
Wakimoto formulas. Let us consider the Heisenberg algebra $\tilde \Gamma$
with generators $a_1(n),{\mathop a\limits^+}_1(n), a^1(n),
{\mathop a\limits^+}^1(n), \rho(n)=\oint_0 dzz^ni\pa \rho (z),
\lambda (n)=\oint_0dzz^ni\pa \lambda (z)$. Let $F_{(l,r)}$ be
irreducible representation of $\tilde \Gamma$ with vacuum vector
$\vartheta_{l,r}$ annihilated by $\ap^1(n),\ap_1(n),\rho(n), \lambda (n), n>0$
and by $a^1(n),a_1(n),n\ge 0$, such that
\eq
\rho(0)\vartheta
_{l,r}=&{2r\over{\sqrt{k+2}}}\vartheta
_{l,r};\  \lambda (0)={2l\over{\sqrt{k+2}}}
\vartheta
_{l,r}&(16)\cr
\endeq
The action of the generators $R_E(n)={\mathop\oint\limits_0}
dzz^nR_E(z)$,
$L_F(n)={\mathop\oint\limits_0}
dzz^nL_F(z)$
on $\vartheta_{l,r}$ is defined if
\eq
&2(r+l)=N(k+2),\qbox{where}
N\in\Z&(17)\cr
\endeq
In this case the generators $R_E(n)$,
$L_F(n)$ acts from $F_{(l,r)}$
to another $F_{(l',r')}$, such that $l'+r'=l+r$.
It is natural to consider the direct sum of Fock modules
\eq
&M_N=\bigoplus_{r,l\atop 2(r+l)=N(k+2)}
F_{(r,l)}&(18)\cr
\endeq
as a representation of $\widehat{\sl(2,\C)}\oplus \widehat{\sl(2,\C)}$.
Let $\g_b$ be subalgebra in the left
$\widehat{\sl(2,\C)}$ which consists in $\{L_E(n),n\in\Z, L_H(n),
n>0\}$.
The algebra $\g_b$ has a natural decomposition
$\g_b=\g^+_b\oplus\g^-_b$,
$\g^+_b=\{L_E(n),n>0,L_H(n),n>0\}$,
$\g^-_b=\{L_E(n),n\le 0\}$.
This decomposition gives
us possibility to define the semi-infinite cohomology
of $\g_b$ with coefficients in
$M_N$ [6].
On the cohomology the right $\widehat{\sl(2,\C)}$
is acting and this is given exactly by the Wakimoto formulas.
Remainder of the left action is the screening operator.
\par
The $\widehat{\sl(3,\C)}\oplus \widehat{\sl(3,\C)}$
currents are given by
\eq
L_{E_1}=&a_1\cr
R_{E_1}=&-:\ap^1(\ap^1a^1-\ap^2a^2+\ap^3a^3)
:-\ap^3a^2-i\sqrt{k+3}\ap^1
\pa\rho_1\cr
&-(k+6)\pa\ap^1+\exp\left[
-{i\over\sqrt{k+3}}(\rho_1+\lambda_1)\right]
(a_1+\ap_2a_3)\cr
L_{E_2}=&a_2+\ap_1a_3\cr
R_{E_2}=&-:(\ap^2)^2a^2:+\ap^3a^1-
i\sqrt{k+3}\ap^2\pa\rho_2-(k+5)\pa\ap^2
+\exp\left[-{i\over\sqrt{k+3}}(\rho_2+\lambda_2)\right]
a_2\cr
L_{H_1}=&-2:\ap_1a_1:+:\ap_2a_2:-:\ap_3a_3:+i\sqrt{k+3}
\pa\lambda_1\cr
R_{H_1}=&2:\ap^1a^1:-:\ap^2a^2:+:\ap^3a^3:+:i\sqrt{k+3}\pa\rho_1&(19)\cr
L_{H_2}=&:\ap_1a_1:-2:\ap_2a_2:-:\ap_3a_3:+i\sqrt{k+3}\pa\lambda_2\cr
R_{H_2}=&-:\ap^1a^1:+2:\ap^2a^2:+:\ap^3a^3:+i\sqrt{k+3}\pa\rho_2\cr
L_{F_2}=&-:(\ap_2)^2a_2:+\ap_3a_1+
i\sqrt{k+3}\ap_2\pa\lambda_2
+(k+1)\pa\ap_2+\exp\left[-{i\over\sqrt{k+3}}
(\rho_2+\lambda_2)\right]a^2\cr
R_{F_2}=&a^2+\ap^1a^3\cr
L_{F_1}=&-:\ap_1(\ap_1a_1-\ap_2a_2+\ap_3a_3):-
\ap_3a_2+i\sqrt{k+3}\ap_1\pa\lambda_1\cr
&+k\pa\ap_1+\exp\left[-{i\over\sqrt{k+3}}(\rho_1+\lambda_1)\right]
(a^1+\ap^2a^3)\cr
R_{F_1}=&a^1.\cr
\endeq
\beginsection
\S 4. Regular representations of $\widehat{sl(n+1)}$ Kac-Moody algebras

The generalization of (15), (19) for $\widehat{sl(n+1,\C)}$ algebras is
immediate. Let $\al_1,\cdots ,\al_n$ be the set of simple roots of
$sl(n+1,\C)$. Denote
\eq
\ap_{ij}(z)=&\ap_{(\al_i+\cdots +\al_j)}(z),\
a_{ij}(z)=a_{(\al_i+\cdots +\al_j)}(z),\  l\le i\le j\le n\cr
\ap^{ij}(z)=&\ap_{-(\al_i+\cdots +\al_j)}(z),\
a^{ij}(z)=a_{(\al_i+\cdots +\al_j)}(z),\  1\le i\le j\le n&(20)\cr
\nbox{and put:}
a_{ij}(z)\ap_{nm}(w)=&a^{ij}(z)\ap^{nm}(w)=(z-w)^{-1}\delta_{in}\delta_{jm}
+\cdots&(21)\cr
\endeq
Introduce the set of free bosonic fields $\la_i(z), \rho_i(z),i=1,\cdots, n$
and put:
\eq
-\la_i(z)\la_j(w)=&\rho_i(z)\rho_j(w)=ln(z-w)K_{ij},&(22)\cr
\endeq
where $K_{ij}$ is Cartan matrix of $sl(n+1)$. Denote $\nu^2=k+n+1$.
Then
\eq
L_{E_{n+1-i}}=&a_{n+1-i,n+1-i}+\sum^n_{j=i+1}:\ap_{n+1-j,n-i}
a_{n+1-j,n+1-i}:\cr
R_{E_{n+1-i}}=&:\ap^{n+1-i,n+1-i}(\sum^{j-1}_{j=1}\ap^{n+2-i,n+1-j}
a^{n+2-i,n+1-j}\cr
&-\sum^i_{j=1}\ap^{n+1-i,n+1-j}a^{n+1-i,n+1-j}):\cr
&-i\nu \ap^{n+1-i,n+1-i}\partial \rho_{n+1-i}+\sum^n_{j=i+1}:
\ap^{n+1-j,n+1-i}a^{n+1-j,n-i}:\cr
&-\sum^{i-1}_{j=1}:\ap^{n+1-i,n+1-j}a^{n+2-i,n+1-j}:-(\nu^2+1+i)\partial
\ap^{n+1-i,n+1-i}\cr
&+\exp [{-i\over \nu}(\rho +\la)_{n+1-i}](a_{n+1-i,n+1-i}
+\sum^{i-1}_{j=1}:\ap_{n+2-i,n+1-j}a_{n+1-i,n+1-j}:)\cr
L_{H_{n+1-i}}=&-2:\ap _{n+1-i,n+1-i}a_{n+1-i,n+1-i}:\cr
&-\sum^{i-1}_{j=1}:(\ap_{n+1-i,n+1-j}a_{n+1-i,n+1-j}
-\ap_{n+1-j,n+2-i}a_{n+1-j,n+2-i}):\cr
&-\sum^n_{j=i+1}:(\ap_{n+1-i,n+1-i}
a_{n+1-j,n+1-i}-\ap_{n+1-j,n-i}a_{n+1-j,n-i}):\cr
&+i\nu \partial \la_{n+1-i}&(23)\cr
R_{H_{n+1-i}}=&2:\ap^{n+1-i,n+1-i}a^{n+1-i,n+1-i}:\cr
&+\sum^{i-1}_{j=1}:(\ap^{n+1-i,n+1-j}a^{n+1-i,n+1-j}-\ap^{n+1-j,n+2-i}
a^{n+1-j,n+2-i}):\cr
&-\sum^n_{j=i+1}:(\ap^{n+1-j,n+1-i}a^{n+1-j,n+1-i}-\ap^{n+1-j,n-i}
a^{n+1-j,n-i})\cr
&+i\nu \partial \rho_{n+1-i}\cr
L_{F_{n+1-i}}=&:\ap_{n+1-i,n+1-i}(\sum^{i-1}_{j=1}\ap_{n+2-i,n+1-j}
a_{n+2-i,n+1-j}\cr
&-\sum^i_{j=1}\ap_{n+1-i,n+1-j}a_{n+1-i,n+1-j}):
+i\nu\ap_{n+1-i,n+1-i}\pa\la_{n+1-i}\cr
&+\sum^n_{j=i+1}:\ap_{n+1-j,n+1-i}a_{n+1-j,n-i}:\cr
&-\sum^{i-1}_{j=1}:\ap_{n+1-i,n+1-j}a_{n+2-i,n+1-j}:
+(\nu^2-1-i)\partial \ap_{n+1-i,n+1-i}\cr
&+\exp [-{i\over \nu}(\rho +\la)_{n+i-1}](a^{n+1-i,n+1-i}
+\sum^{i-1}_{j=1}:\ap^{n+2-i,n+1-j}a^{n+1-i,n+1-j}:)\cr
R_{F_{n+1-i}}=&a^{n+1-i,n+1-i}+\sum^n_{j=i+1}:\ap^{n+1-j,n-i}
a^{n+1-j,n+1-i}:\cr
\endeq
determine the structure of regular representation of the
$\widehat{sl(n+1,\C)}\oplus \widehat{sl(n+1,\C)}$ affine Kac-Moody algebra.

Note that the structure of these formulas is the following. We have
two copies of free fields: $\{ a_{ij},\ap_{ij}, \la_i\} $ and
$\{ a^{ij},\ap^{ij},\rho _i\} $. Then we write down Wakimoto formulas
for the action of left and right $\widehat{sl(n+1,\C)}$ in terms of
these free fields. The next step - we add to the action of $F_i$ from the
left algebra the ``screening'' currents for the right algebra. And we
also add left ``screening'' currents to the action of $E_i$ from right
algebra. The similar procedure can be done for arbitrary semi-simple Lie
algebra.
\beginsection
References

\item{1.} Wakimoto, M., {\it Commun. Math. Phys.}, {\bf 104} (1986), 605.

\item{2.} Feigin, B., Frenkel, E., {\it Russ. Math. Surveys}, {\bf 43},
N5 (1988), 221-222; {\it Commun. Math. Phys.}, {\bf 128} (1990), 161-189.

\item{3.} Bernard, D., Felder, G., {\it Commun. Math. Phys.}, {\bf 127}
(1990), 145-168.

\item{4.} Feigin, B., Frenkel, E., in V. Kniznik Memorial Volume,
eds. L. Brink, e.a., 271-316, World Scientific, Singapore, 1989.

\item{5.} Bershadsky, M., Ooguri, H., {\it Comm. Math. Phys.}, {\bf 126}
(1989), 49.

\item{6.} Feigin, B., Semi-infinite homology of Kac-Moody and Virasoro Lie
algebras, {\it Usp. Mat. Nauk (=Russ. Math. Surv.)}, {\bf 39} (1984),
195-196 (in Russian).

\end